\documentclass{llncs}

\usepackage[utf8]{inputenc}

\usepackage{graphics} 
\usepackage{epsfig} 
\usepackage{mathptmx} 
\usepackage{times} 
\usepackage{amsmath} 
\usepackage{amssymb}  
\usepackage{multirow}
\usepackage{lineno}
\usepackage{tikzpagenodes}
\usepackage{multirow}
\usepackage{bigdelim}

\usepackage[dvips]{rotating} 
\usepackage{array}
\usepackage{multirow}
\usepackage{subfigure}
\usepackage{rotating}

\usepackage{textcomp}
\usepackage{epsfig}

\author{**********\\*************}
\institute{**********\\************}
\title{Improving Endoscopic Decision Support Systems by  Translating Between Imaging Modalities}
\author{Georg Wimmer~\inst{1,*}, Michael Gadermayr~\inst{2,*}, Andreas V\'ecsei~\inst{3}, Andreas Uhl~\inst{1}}

\institute{
$^1$ University of Salzburg, Department of Computer Science, Salzburg, Austria\\
$^2$ Salzburg University of Applied Sciences, Salzburg, Austria\\
$^3$ St. Anna Children's Hospital, Vienna, Austria, $^*$ Equal Contributions
}

\begin{document}

\maketitle

\begin{abstract}
Novel imaging technologies raise many questions concerning the adaptation of computer-aided decision support systems. Classification models either need to be adapted or even newly trained from scratch to exploit the full potential of enhanced techniques. Both options typically require the acquisition of new labeled training data.
In this work we investigate the applicability of image-to-image translation to endoscopic images showing different imaging  modalities, namely conventional white-light and narrow-band imaging.
In a study on computer-aided celiac disease diagnosis, we explore whether image-to-image translation is capable of effectively performing the translation between the domains. We investigate if models can be trained on virtual (or a mixture of virtual and real) samples to improve overall accuracy in a setting with limited labeled training data. Finally, we also ask whether a translation of testing images to another domain is capable of improving accuracy by exploiting the enhanced imaging characteristics.
\end{abstract}


\section{Motivation}\label{sec:introduction}
Optimum visibility of the crucial visual features for a reliable diagnosis is an important criterion for imaging devices in endoscopy.
%
%
Narrow-band imaging (NBI)~\cite{Emura08a,Valitutti14a} is an imaging technique, where light of specific blue and green wavelengths is utilized to enhance the detail of the surface of the mucosa. A filter is electronically activated leading to the use of ambient light of wavelengths of 415 (blue) and 540 nm (green) only. Conventional endoscopes illuminate the mucosa with a broad visible light-spectrum.
%
Strong indication is provided that also computer-aided decision support systems show improved accuracy if being applied to NBI data instead of conventional white-light imaging (WLI) data~\cite{Wimmer19a}. 
%
%
Due to the clearly different image characteristics of the two imaging modalities NBI and WLI, classification models trained for one of the modalities cannot be directly applied to the other modality without clearly loosing accuracy~\cite{Wimmer19a}. The straight-forward approach is to train an individual model for NBI and one for WLI. However, convolutional neural networks (CNNs), which are the state-of-the-art in endoscopic image classification, require large amounts of labeled data showing high visual quality. As there is typically a lack of such data, a major question is, if data from both domains can be used to train either individual models or one generic model for both domains. For that purpose, typically domain adaptation methods are applied.
Conventional domain adaptation integrates the adaptation of the data in the image analysis pipeline. 
For that reason, domain adaptation is typically linked to the final image analysis approach. Image-to-image translation can be interpreted as domain adaptation on image level~\cite{myKamnitsas16a}
and  is therefore completely independent of the final task as only the image itself is converted. 
Image-to-image translation gained popularity during the last years generating highly attractive and realistic output~\cite{myIsola16a,myZhu17a}. 
While many approaches require image pairs for training the image transformation models, generative adversarial networks (GANs)~\cite{myIsola16a}, making use of the so-called cycle-consistency loss (cycle-GANs), are free from this restriction. 
They do not require any image pairs for domain adaptation and are completely independent of the underlying image analysis (e.g. classification) model. Finally, they also improve interpretability of the adaptation process by allowing visual investigation of the generated virtual (or "fake") images.
For the purpose of domain adapatation, cycle-GANs were effectively used in the field digital pathology~\cite{Gadermayr19b} and the field of radiology for translating between the modalities, such as MRI and CT~\cite{myWolterink17a}.

Celiac disease (CD) is a multisystemic immune-mediated disease, which is associated with considerable morbidity and mortality~\cite{Biagi10a}.
%
Endoscopy in combination with biopsy is currently considered the gold standard for the diagnosis of CD. 
Computer-aided decision support systems have potential to improve the whole diagnostic work-up, by saving costs, time and manpower and at the same time increase the safety of the procedure~\cite{Wimmer19a}.
A motivation for such a system is furthermore given as the inter-observer variability is reported to be high \cite{Biagi10a,Biagi06a}. 
%

\subsection{Contributions} 
In this work, we investigate whether decision support systems for celiac disease diagnosis can be improved by making use of image-to-image translation. Specifically, we perform translations between images captured using the NBI technology and conventional WLI. Several experiments are performed in order to answer the following questions: (Q1) Can image-to-image translation be effectively applied to endoscopic images? (Q2) Can virtual images be used to increase the training data set and finally improve classification accuracy? (Q3)  And how do virtual images perform when being used during testing? Do fake images in the more discriminative NBI domain maybe even show improved performance compared to real WLI samples?
For answering these questions, several pipelines (Sect.~\ref{sec:setup}) combining classification models~(Sect.~2.2) with an image translation model~(Sect.~2.1) are developed and evaluated.

\section{Methods}


\subsection{Image Translation Network} \label{sec:i2i}

For performing translation between the NBI and the WLI data (and vice versa), we employ a generative adversarial network (GAN). Specifically we utilize the so-called cycle-GAN model~\cite{myZhu17a} which allows unpaired training. Unpaired training in this setting means that training only requires one data set for each of the two domains. Corresponding images, which would be hardly impossible to obtain in our application scenario, are not needed for this approach.

Cycle-GAN consists of two generator models, $F:{X} \rightarrow {Y}$ and $G:{Y} \rightarrow {X}$ and two discriminators $D_X$ and $D_Y$ which are trained optimizing the cycle-consistency loss
\begin{equation}\begin{split}
\mathcal{L}_{c}=\mathbb{E}_{x \sim p_{data}(x)} [|| G(F({x})) - {x} ||_1 ] +  
\mathbb{E}_{y \sim p_{data}(y)} [ ||F(G({y})) - {y}||_1 ]
\end{split}
\end{equation}
as well as the adversarial loss (or GAN loss)
\begin{equation}\begin{split}
\mathcal{L}_{d} =
\mathbb{E}_{x \sim p_{data}(x)} [\log (D_X(x)) \hspace{-0.03cm} + \hspace{-0.03cm} \log (1 \hspace{-0.03cm} - \hspace{-0.03cm} D_Y(F(x)))] \; + \\
\mathbb{E}_{y \sim p_{data}(y)} [\log (1-D_X(G(y))) + \log (D_Y(y))] \; .
\end{split}\end{equation}
$F$ and $G$ are trained to fool the discriminators and thereby generate fake images that look similar to real images, while $D_X$ and $D_Y$ are optimized to distinguish between translated and real samples for both domains individually.
While the generators aim to minimize this adversarial loss, the discriminators aim
to maximize it.

To train the image-translation model, for both data sets, all original endoscopic images are padded to fit a common size (768 $\times$ 768).
With these patches, a cycle-GAN is trained~\cite{myZhu17a} (with equal weights $w_{cyc}=1, w_{GAN} = 1$).
We do not use the optional identity loss $\mathcal{L}_{id}$.
As generator network, we employ the U-Net~\cite{myRonneberger15a}.
Apart from that, the standard configuration based on the patch-wise CNN discriminator is utilized~\cite{myZhu17a})\footnote{We use the provided PyTorch reference implementation~\cite{myZhu17a}.}.
Training is performed for 1000 epochs. The initial learning rate is set to $10^{-5}$. Random flipping and rotations (0, 90, 180, 270$^\circ$) are applied for data augmentation.

\subsection{Classification Models} \label{sec:classification}

We investigate three CNNs for performing classification of celiac disease based on endoscopic image patches. The networks consist of AlexNet~\cite{Krizhevsky2012a}, VGG-f~\cite{Chatfield14a} and VGG-16~\cite{Simonyan15a}. 
These models are chosen as they exhibit high potential for endoscopic image analysis~\cite{Wimmer19a,myRibeiro16a}.
The networks are trained on manually selected patches of a size of $256 \times 256$ pixels.
The selection is performed because major image ares do not show any discriminative features for distinguishing between the classes~\cite{Wimmer19a}.
To prevent any bias, all networks are randomly initialized and trained from scratch. 
Due to the small available data set and the accompanying problems during training~\cite{myRibeiro16a}, we did not employ deeper architectures such as ResNet or Inception-Net.
The size of the last fully-connected layer is adapted to the considered two-classes classification scheme. 
The last fully connected layer is acting as soft-max classifier and computes the training loss (log-loss).
Stochastic gradient descent with weight decay $\lambda = 0.0005$ and momentum $\mu = 0.9$ is employed for the optimization.
Training is performed on batches of 128 images each, which are randomly selected (independently for each iteration) from the training data and subsequently augmented. For augmentation, random cropping ($224 \times 224$ pixel patches for VGG-f and VGG-16 and $227 \times227$ pixel patches for Alex-net), horizontal flipping and random rotations with multiples of 90\textdegree are performed. 

\begin{figure}[bt]
	\centering
	\subfigure[NBI, healthy]{\includegraphics[width=0.235\columnwidth]{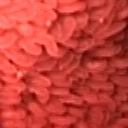} \includegraphics[width=0.235\columnwidth]{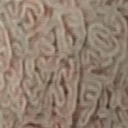}} $\;\;\;$
	\subfigure[NBI, CD]{\includegraphics[width=0.235\columnwidth]{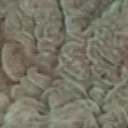} 
	\includegraphics[width=0.235\columnwidth]{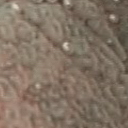}}
	\subfigure[WLI, healthy]{\includegraphics[width=0.235\columnwidth]{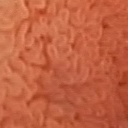} 
	\includegraphics[width=0.235\columnwidth]{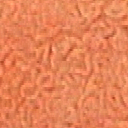}} $\;\;\;$
	\subfigure[WLI, CD]{\includegraphics[width=0.235\columnwidth]{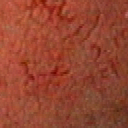} 		\includegraphics[width=0.235\columnwidth]{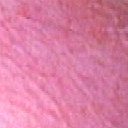}}\\
	
	\caption{Example images for the two classes healthy and celiac disease (CD) using NBI as well as WLI endoscopy.}
	\label{fig:marsh}
\end{figure}


\subsection{Experimental Setup} \label{sec:setup}

In this work we consider a two-classes classification and differentiate between healthy mucosa and mucosa affected by CD using images gathered by NBI as well as WLI (both in combination with the modified immersion technique~\cite{Wimmer19a}).
The WLI image database consists of 1045 image patches (587 healthy images and 458 affected by CD) and the NBI database consists of 616 patches  (399 healthy images and 217 affected by celiac disease). 
The RGB image patches of both data sets measure $256 \times 256$ pixels. They are gathered by manually extracting image patches from endoscopic image material showing distinctive regions-of-interest.
Figure~\ref{fig:marsh} shows example patches showing healthy mucosa and mucosa affected by CD, for both considered imaging modalities individually.

In all experiments, five-fold cross-validation is performed to achieve a stable estimation of the generalization error, where each of the five folds consists of the images of approx. 20\% of all patients.
To avoid bias, we ensure that all images of one patient are in the same fold.
By means of the McNemar test, 
we assess the statistical significance of the obtained improvements.

To be able to answer the raised questions Q1, Q2 and Q3, we not only perform experiments (E) using training and evaluation data showing the same imaging modality (referred to as E1), but we also perform experiments using domain transferred (virtual or fake) samples and (only for training) also mixtures of domain transferred and original samples.

To answer \textbf{Q2} ("Can domain transferred images be used to increase the training data set and finally improve classification accuracy?"), we perform the following experiments (E2):
Firstly, training and testing with original images is performed separately for WLI and NBI. Secondly, NBI and WLI images are combined for training and a generic model that is applied to both data sets for testing. Finally, NBI and fake NBI (NBI$_f$) samples (i.e. WLI samples which were translated to NBI) are used to train a model for NBI. The same is applied to WLI samples using real WLI and fake WLI (WLI$_f$) samples to train a model to classify WLI data.
In all settings, we take care that samples of each patient are either in the training set or in the test set (also when combining NBI and WLI images).

To answer \textbf{Q3} ("How do domain transferred images perform when being used during testing?"), the following experiments (E3) are performed:
First, we compare the outcomes for using either original image data for training and evaluation or for using domain transferred image data for training and evaluation.
In a second experiment, real and virtual (either real NBI and virtual WLI or vice versa) samples are combined for training to enlarge the training corpus and classification is performed either on original or virtual image samples showing the same real or fake domain.
In that way, we want to find out if the virtual data is better suited for classification than the original endoscopic image data. This is especially promising in the case of transforming WLI to NBI, since NBI is supposed to improve the results of automated diagnosis systems compared to WLI~\cite{Wimmer19a}. 

\textbf{Q1} ("Does image-to-image translation work for endoscopic images?") is finally answered by means of visual inspection of the generated domain transferred images and by considering the quantitative experiments mentioned above.

\section{Results} \label{sec:results}

In Table~\ref{table:result}, the overall classification rates for the five performed experiments are shown.
For each CNN architecture and each Experiment, the result of the best performing combination of training and test data set is given in boldface numbers.
The scores within E2a, E2b, E3a and E3b can be directly compared as the underlying image content is the same for considered testing data. A direct unbiased numeric comparison in case of E1 and between experiments (e.g. between E2a and E2b) is not possible since the data sets show different mucosal areas. Consequently, E1 does not directly answer the question for the more suitable imaging modality for computer aided diagnosis.

\begin{table*}[tb]
	\begin{center} 
		\begin{tabular}{c c c c c c c c c }
			
			$\;\;$Training$\;\;$	& $\;\;\;\;\;$Test$\;\;\;\;\;$ 	& \multicolumn{2}{l}{$\;\;$VGG-f$\;\;\;\;\;\;\;$}	& \multicolumn{2}{l}{$\;\;$Alex-net$\;\;\;\;\;$}	&\multicolumn{2}{l}{$\;\;$VGG-16$\;\;\;\;\;$} & $\varnothing$\\ \hline
			\vspace{-0.1cm} \\
			\multicolumn{9}{c} {E1: Baseline for WLI and NBI data}\\ \hline
			WLI		&		WLI		&
			87.4(2.4)	& 	&	87.6(1.5) &	&	85.9(2.5)	&	&	87.0\\
			NBI		&		NBI		&
			87.6(3.8) &		&	88.1(4.1) &	& 83.5(3.4)	&		&	86.4\\\hline \vspace{-0.1cm} \\
			\multicolumn{9}{c} {E2a: Accuracies for testing NBI samples}\\
			\hline
			NBI					&NBI		&87.6(3.8)	 & 		&\textbf{88.1}(4.1)	&&83.5(3.4)				&&86.4\\
			NBI $\cup$ WLI		&NBI		&81.3(3.9)			&&81.5(6.0)			&&80.3(5.3)				&&81.0\\
			NBI $\cup$ NBI$_f$	&NBI		&\textbf{88.3}(2.6)	&&87.0(2.4)			&&\textbf{84.9}(5.3)		&&\textbf{86.7}\\ \hline
			\vspace{-0.1cm} \\
			\multicolumn{9}{c} {E2b: Accuracies for testing WLI samples}	\\
			\hline
			WLI		&WLI		&87.4(2.4)& \hspace{-0.4cm} \multirow{3}{0.1cm}{\bigg]*}	&87.6(1.5)& \hspace{-0.4cm} \multirow{3}{0.1cm}{\bigg]*}	&85.9(2.5)&		&87.0\\
			NBI $\cup$ WLI		&WLI		&87.3(2.4)&	&88.2(1.3)&	&86.3(2.2)&		&87.3\\
			WLI$_f$ $\cup$ WLI	&WLI		&\textbf{89.8}(1.6)& &\textbf{89.6}(1.7)& 	&\textbf{87.5}(3.3)&	&\textbf{89.0}\\ \hline
			\vspace{-0.1cm} \\
			\multicolumn{9}{c} {E3a: Converting WLI to NBI for testing}\\ \hline
			WLI		&WLI		&87.4(2.4)& \hspace{-0.4cm} \multirow{2}{0.1cm}{\big]*}	 &87.6(1.5)& \hspace{-0.4cm} \multirow{2}{0.1cm}{\big]*}  &85.9(2.5)&		&87.0\\
			WLI$_f$ $\cup$ WLI	&WLI		&\textbf{89.8}(1.6)& 	&\textbf{89.6}(1.7)&  &87.5(3.3)&	&\textbf{89.0}\\
			NBI$_f$		&NBI$_f$	&88.9(2.2)&	&89.5(2.0)&	&\textbf{88.2}(3.1)&		&88.9\\
			NBI $\cup$ NBI$_f$	&NBI$_f$	&88.8(2.5)&	&89.2(1.4)&	&86.6(2.4)&		&88.2\\	\hline
			\vspace{-0.1cm} \\
			\multicolumn{9}{c} {E3b:  Converting NBI to WLI for testing}\\ \hline
			NBI		&NBI		&87.6(3.8) &	&\textbf{88.1}(4.1)	&&83.5(3.4)	&	&86.4\\
			NBI $\cup$ NBI$_f$	&NBI	&\textbf{88.3}(2.6)&&87.0(2.4)&	&\textbf{84.9}(5.3)&		&\textbf{86.7}\\
			WLI$_f$		&WLI$_f$	&82.9(2.9) & &81.9(2.8) & &78.5(2.7)	& 	&81.1\\
			WLI $\cup$ WLI$_f$	&WLI$_f$	&85.6(5.6) &	&85.1(6.5) &	&83.5(6.7)&		&84.7\\	\hline
			
		\end{tabular} 
	\end{center}
	\caption{Mean classification accuracies  for the CNNs using different combinations of training and test data sets. WLI: white-light endoscopic data; NBI: NBI endoscopic data; WLI$_f$: NBI endoscopic data domain transformed to white-light; NBI$_f$: white-light data domain transfered to NBI. The asterisk (*) indicates statistical significant improvements ($p<0.05$).}
	\label{table:result}
\end{table*}

When comparing the outcomes of the standard setting with the same domain for training and testing (E1), we notice that the accuracies are slightly higher (+0.6\% on average) in case of WLI image data.  
However, here we need to remark that the overall data set (and hence also the data available for training) is larger (WLI: 1045 patches; NBI: 616 patches).
Additionally, the testing data does not capture the same mucosal regions.

In E2a and E2b, the effect of fake training data is assessed (Q2).
Using only NBI images as training data for testing NBI images (E2a) showed higher scores (+5.4\% on average) than mixing NBI and WLI data (without any domain adaptation) for training. 
Combining NBI data with virtual NBI$_f$ samples for training on the other hand slightly improved the results (+0.3\% on average) compared to using NBI data only for training.

\begin{figure}[tb] \center
	\includegraphics[width=0.99\linewidth]{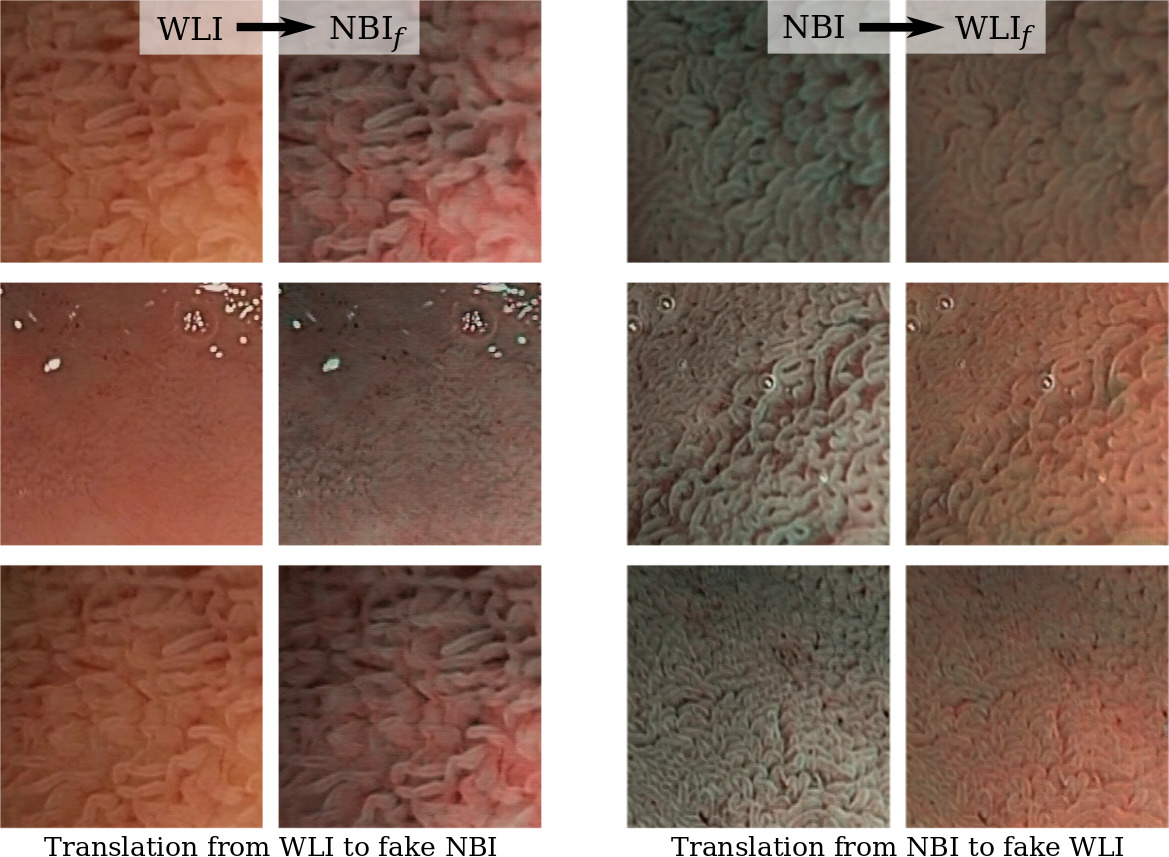}
	\caption{
		Example images before and after  the image-to-image translation process. The figures on the left  show original WLI samples and the original WLI samples  translated to NBI (NBI$_f$) and the figures to the right show original NBI samples and the original NBI samples translated to WLI  (WLI$_f$).} 
	\label{fig:translation}
\end{figure}

Using only WLI data for training when testing on WLI images (E2b) achieved the lowest scores. Additionally using NBI data for training, slightly improved the accuracies (+0.3\% on average) but the clearly best results were achieved combining WLI images and virtual WLI$_f$ images (+2.0\% on average).


In E3a and E3b, the effect of translating samples for testing is assessed (Q3). 
Considering the WLI samples for testing (E3a), the best scores were achieved when testing the original WLI data using WLI and virtual WLI$_f$ data for training. 
When training and testing on WLI data converted to NBI (NBI$_f)$) showed similar rates (-0.1\% on average).

Considering the NBI samples for testing, the best accuracies were achieved when testing the original NBI data (83.5-88.3\% on average). When classifying NBI data converted to WLI, accuracies drop to between 78.5 and 85.6\%. 

Example translated "fake" images and the corresponding original "real" samples are provided in Fig.~\ref{fig:translation} for both translation directions.

\section{Discussion}
In this paper, we investigate whether decision support systems for celiac disease diagnosis can be improved by making use of image translation. Experiments were performed to answer three questions posed in Sect.~\ref{sec:introduction}. 

We first discuss whether image-to-image translation in principle works for the considered endoscopic image data (Q1)? Regarding the example fake images and the quantitative final classification scores, we can conclude that image translation is able to generate useful fake images in both directions. In general the images look realistic. In some cases, we notice that the translation process generates images with a high similarity to the original domain. This is supposed to be due to the fact that the data sets also contain images which look like hybrids between NBI and WLI depending on the illumination quality and the perspective of the camera towards the mucosal wall. There are huge differences with respect to illumination and perspective in the endoscopic image data. 
Regarding the overall classification rates, however, we notice that domain transferred images, either used alone or in combination with original images, generally show high accuracy (especially WLI data converted to NBI).

Next, we asked whether virtual images should be used to enlarge the training data set to finally improve classification accuracy in settings with limited annotated training data (Q2). 
We can summarize that this is definitely the case. We notice increased accuracies in five out of six settings when combining fake and original samples for classification. This effect is observed independent of the direction of translation, i.e. improvements are obtained for NBI-to-WLI translation as well as for WLI-to-NBI translation.
The accuracies also clearly increase when combining e.g. NBI and NBI$_f$ for training instead of simply combining the NBI and the WLI data set when testing NBI samples.
For that reason, we can conclude that the effect of the domain shift between NBI and WLI data is clearly reduced when performing image translation. 

Finally, we asked how domain transferred images perform when being used during testing (Q3)? To answer this question, a differentiation between the two data sets needs to be done. A translation of test data to the WLI domain (WL$_f$) generally shows decreased rates. As already indicated in~\cite{Wimmer19a}, this is probably because the WLI modality is generally less suited for a computer-based classification than NBI. Even a perfect translation to a less suited modality obviously cannot exhibit better scores.
Considering a translation to the NBI domain, we notice improvement if fake samples are used for training and testing only. This also corroborates the assumption that NBI can be more effectively classified and that the translation process works really well.
Interestingly, using original samples additionally to the domain transferred ones does not show further improvements. Even though the translation in principle works well, it can be assumed that there is still some domain shift left between real and fake samples. Finally, it could be the case that the virtual NBI$_f$ domain requires fewer samples for effectively training a classification network. This could be because of clearer highlighted features and/or due to decreased variability in the data.

\section{Conclusion}

We showed that image-to-image translation can be effectively applied to endoscopic images in order to change the imaging modality from WLI to NBI and vice versa. We showed that additionally employing virtual training data does increase the classification rates of CNNs.
Converted images applied for testing also showed slightly increased accuracies if conversion is performed from WLI to the obviously more appropriate NBI domain. 
This study therefore also provides (further) indication that NBI is more suited than WLI for computer aided decision-support systems since the translation to the NBI domain clearly achieves better results than the translation to WLI.
We assume that the slightly higher accuracies for WLI in the standard setting (with unprocessed data for training and testing) are mainly caused by the larger training data.
Finally, the obtained insights provide incentive to attempt to further increase distinctiveness by translating to special subdomains showing particularly high image quality. A distinctiveness-contraint could also be incorporated into the GAN architecture.
%


\newpage

\bibliographystyle{splncs04}
\bibliography{bibsNBI,my,eigene,all}

\begin{thebibliography}{10}
\providecommand{\url}[1]{\texttt{#1}}
\providecommand{\urlprefix}{URL }
\providecommand{\doi}[1]{https://doi.org/#1}

\bibitem{Biagi10a}
Biagi, F., Corazza, G.: Mortality in celiac disease. Nature Reviews
  Gastroenterology and Hepatology  \textbf{7},  158--162 (2010)

\bibitem{Biagi06a}
Biagi, F., Rondonotti, E., Campanella, J., Villa, F., Bianchi, P.I., Klersy,
  C., Franchis, R.D., Corazza, G.R.: Video capsule endoscopy and histology for
  small-bowel mucosa evaluation: A comparison performed by blinded observers.
  Clinical Gastroenterology and Hepatology  \textbf{4}(8),  998--1003 (2006)

\bibitem{Chatfield14a}
Chatfield, K., Simonyan, K., Vedaldi, A., Zisserman, A.: Return of the devil in
  the details: Delving deep into convolutional nets. In: British Machine Vision
  Conference, {BMVC} 2014, Nottingham, UK, September 1-5 (2014)

\bibitem{Emura08a}
Emura, F., Saito, Y., H., I.: Narrow-band imaging optical chromocolonoscopy:
  advantages and limitations. World J Gastroenterol  \textbf{14}(31),
  4867--4872 (2008)

\bibitem{Gadermayr19b}
Gadermayr, M., Gupta, L., Appel, V., Boor, P., Klinkhammer, B.M., Merhof, D.:
  Generative adversarial networks for facilitating stain-independent supervised
  \& unsupervised segmentation: A study on kidney histology. IEEE Transactions
  on Medical Imaging  (2019)

\bibitem{myIsola16a}
Isola, P., Zhu, J.Y., Zhou, T., Efros, A.A.: Image-to-image translation with
  conditional adversarial networks. In: Proceedings of the International
  Conference on Computer Vision and Pattern Recognition (CVPR'17) (2017)

\bibitem{myKamnitsas16a}
Kamnitsas, K., Baumgartner, C.F., Ledig, C., Newcombe, V.F.J., Simpson, J.P.,
  Kane, A.D., Menon, D.K., Nori, A.V., Criminisi, A., Rueckert, D., Glocker,
  B.: Unsupervised domain adaptation in brain lesion segmentation with
  adversarial networks. CoRR, http://arxiv.org/abs/1612.08894  (2016)

\bibitem{Krizhevsky2012a}
Krizhevsky, A., Sutskever, I., Hinton, G.E.: Imagenet classification with deep
  convolutional neural networks. In: Advances in Neural Information Processing
  Systems 25, pp. 1097--1105. Curran Associates, Inc. (2012)

\bibitem{myRibeiro16a}
Ribeiro, E., Uhl, A., Wimmer, G., H\"{a}fner, M.: Exploring deep learning and
  transfer learning for colonic polyp classification. Computational and
  Mathematical Methods in Medicine  \textbf{2016},  1--16 (2016)

\bibitem{myRonneberger15a}
Ronneberger, O., Fischer, P., Brox, T.: U-net: Convolutional networks for
  biomedical image segmentation. In: Proceedings of the International
  Conference on Medical Image Computing and Computer Aided Interventions
  (MICCAI'15), pp. 234--241 (2015)

\bibitem{Simonyan15a}
Simonyan, K., Zisserman, A.: Very deep convolutional networks for large-scale
  image recognition. In: International Conference on Learning Representations
  (ICLR) (2015)

\bibitem{Valitutti14a}
Valitutti, F., Oliva, S., Iorfida, D., Aloi, M., Gatti, S., Trovato, C.M.,
  Montuori, M., Tiberti, A., Cucchiara, S., Di~Nardo, G.: Narrow band imaging
  combined with water immersion technique in the diagnosis of celiac disease.
  Dig and Liver Dis  \textbf{46}(12),  1099--1102 (2014)

\bibitem{Wimmer19a}
Wimmer, G., Gadermayr, M., Wolkersdoerfer, G., Kwitt, R., Tamaki, T.,
  Tischendorf, J., Haefner, M., Yoshida, S., Tanaka, S., Merhof, D., Uhl, A.:
  Quest for the best endoscopic imaging modality for computer-assisted colonic
  polyp staging. World J Gastroenterol  \textbf{25}(10),  1197--1209 (2019).
  \doi{10.3748/wjg.v25.i10.1197},
  \url{https://www.wjgnet.com/1007-9327/full/v25/i10/1197.htm}

\bibitem{myWolterink17a}
Wolterink, J.M., Dinkla, A.M., Savenije, M.H.F., Seevinck, P.R., van~den Berg,
  C.A.T., I{\v{s}}gum, I.: Deep {MR} to {CT} synthesis using unpaired data. In:
  Proceedings of the International MICCAI Workshop Simulation and Synthesis in
  Medical Imaging (SASHIMI'17). pp. 14--23 (2017)

\bibitem{myZhu17a}
Zhu, J.Y., Park, T., Isola, P., Efros, A.A.: Unpaired image-to-image
  translation using cycle-consistent adversarial networks. In: Proceedings of
  the International Conference on Computer Vision (ICCV'17) (2017)

\end{thebibliography}

\end{document}